\documentclass[11pt]{article}

\usepackage[margin=1in]{geometry} \usepackage{amsmath,amssymb} \usepackage{graphicx} \usepackage{booktabs} \usepackage{hyperref} \usepackage{natbib} \usepackage{caption} \usepackage{subcaption} \usepackage{xcolor}

\title{Laplace-PSN-IRT: Uncertainty Quantification for Neural Item Response Theory Models of LLM Benchmarks}

\author{
Juan Francisco Mandujano Reyes \\ Independent Research \\ \texttt{juan.mandujano@cimat.mx} }

\date{}

\begin{document}
\maketitle

\begin{abstract}
Item Response Theory (IRT) has recently been proposed as a framework for diagnosing the quality of large language model (LLM) benchmarks, separating a model's latent ability from the difficulty, discrimination, and reliability of individual benchmark items. \citet{zhou2026lost} introduce PSN-IRT, as a neural network-based architecture that jointly estimates model ability and a set of item parameters, using point estimates of such parameters (including Fisher information) to diagnose benchmark quality and select smaller, high-quality item subsets. Nevertheless, PSN-IRT and its realted baselines (MLE, MCMC, variational inference, VIBO) report only point estimates. Leaving a gap for credible intervals, posterior-based ranking comparisons, and propagation of parameter uncertainty into downstream item-selection decisions. In this paper, we introduce Laplace-PSN-IRT, a post-hoc last-layer Laplace approximation \citep{mackay1992evidence,daxberger2021laplace} expanding a trained PSN-IRT model, which helps to recover calibrated posterior uncertainty over model ability and item difficulty without retraining the underlying model. Using this posterior, we show that (i) the majority of pairwise comparisons among 12 models on a standard LLM benchmark leaderboard are not statistically distinguishable despite having different point-estimate ranks (only 2 of 66 pairs are confidently separated), (ii) point-estimate Fisher information, the criterion used for efficient item selection, collapses to near-zero for a large, ability-dependent fraction of benchmark items due to saturation at a single reference ability, while the proposed posterior-expected Fisher information remains substantially more stable across the full ability range, and (iii) posterior-expected Fisher information outperforms point-estimate Fisher information at recovering a full-benchmark ability ranking from a small item subset in 8 of 10 of our tested configurations (100-1000 items), while the two criteria are comparable at the smallest subset size tested (50 items). We validate the posterior's calibration directly via held-out predictive coverage and show that treating item difficulty as random while holding item discrimination fixed is necessary for well-calibrated uncertainty in this architecture.
\end{abstract}

\section{Introduction}

Large language model (LLM) benchmarks are usually summarized by a single accuracy metric per model per benchmark, lacking sources of uncertainty: how reliably that metric separates one model from another, and how much of a benchmark's measured ``information'' about model ability is an artifact of a particular evaluation point rather than a robust property of the benchmark itself. Recently, \citet{zhou2026lost} address this by proposing PSN-IRT, a Pseudo-Siamese Network formulation of Item Response Theory (IRT) that estimates, for each evaluated model, a latent ability $\theta$, and for each benchmark item, a difficulty $b$, discrimination $a$, guessing rate $c$, and feasibility ceiling $d$, combined via the following 4-parameter logistic (4PL) model:
\begin{equation}
P(X = 1 \mid \theta, a, b, c, d) = c + (d - c) \cdot \frac{1}{1 + e^{-a(\theta - b)}}.
\label{eq:4pl}
\end{equation}
Using point estimates of these parameters, \citet{zhou2026lost} show that mainstream LLM benchmarks suffer from insufficient difficulty ceilings, item saturation, and likely data contamination. Therefore, they propose selecting selecting a small, high-Fisher-information subset of items to preserve (or improve) agreement with an external human-preference reference ranking (Chatbot Arena), compared to using the full item set.

However, this diagnostic and item-selection pipeline is guided by point estimates. Although, \citeauthor{zhou2026lost} compare against classical Bayesian estimation baselines (MCMC, variational inference, VIBO), these are also evaluated and reported only via point-estimate accuracy and rank-reliability metrics. Therefore, there is gap in the usage of posterior distribution of these methods for calibrated inference. Exploiting Bayesian statistics techniques we can provide reliable diagnostic and item selection by providing credible intervals on model ability, probabilistic statements about which pairs of models are actually distinguishable, and propagation of parameter uncertainty into the Fisher-information item-selection criterion itself \citep{vanderlinden1998bayesian}.

This gap matters specifically for benchmark diagnosis. A leaderboard that reports model $A$ ranked above model $B$ provides different information depending on whether that ordering is almost absolute or susceptible to flip under uncertainty. On the other hand, item-selection criteria computed from a single point estimate of item parameters are, by construction, blind to how much that estimate itself should be trusted. This matters most where the estimate is least trustworthy, such as for items whose parameters are estimated from a small effective sample of contrasting responses.

\paragraph{Contributions.} In this work we introduce Laplace-PSN-IRT, a Bayesian extension of PSN-IRT, which applies a post-hoc last-layer Laplace approximation \citep{mackay1992evidence} to a trained PSN-IRT model. In particular:
\begin{enumerate}
  \item We fit a Gaussian posterior on top of PSN-IRT's trained ability-readout and item-parameter-readout layers using the Laplace approximation \citep{daxberger2021laplace}, without modifying the original architecture or training procedure, and validate its calibration directly against held-out response data. 
  \item We show that this validation surfaces a specific, generalizable failure mode: jointly sampling two parameters (item discrimination $a$ and difficulty $b$) that combine multiplicatively in the model's logit compounds posterior variance beyond what a Gaussian last-layer approximation supports, and that scoping the posterior to difficulty alone (holding discrimination at its point estimate) resolves this, yielding calibration on par with the model-ability posterior.
  \item Using this validated, scoped posterior, we compute posterior-expected Fisher information, $\bar{I}(\theta) = \mathbb{E}_{p(b \mid \mathcal{D})}[I(\theta \mid a, b, c, d)]$, and show it is substantially more stable across the model-ability range than point-estimate Fisher information, and outperforms it as an item-selection criterion for recovering a full-benchmark ability ranking, particularly at subset sizes of 100 items or more.
  \item We report a pairwise, posterior-probability-based comparison of model ability, $P(\theta_i > \theta_j)$, which shows that only a small fraction of pairwise comparisons on a standard 12-model leaderboard are statistically distinguishable, directly quantifying the ``weak separability'' problem raised qualitatively by \citet{zhou2026lost}.
\end{enumerate}

\section{Background: PSN-IRT}

First, we briefly summarize the relevant parts of PSN-IRT \citep{zhou2026lost}; see the original paper for full architectural and training details.

PSN-IRT trains two feed-forward networks on a binary response matrix $X \in \{0,1\}^{M \times N}$ ($M$ models, $N$ items): a \emph{model network} that maps a one-hot model identifier to a latent ability $\theta$, and an \emph{item network} that maps a one-hot item identifier to four item parameters $(a, b, c, d)$ via a shared embedding and a final linear readout layer per branch. These are combined via the 4PL formula (Eq.~\ref{eq:4pl}) and trained end-to-end with binary cross-entropy against observed correctness labels. After training, the two readout layers, \texttt{student\_ability\_out} (model network $\to \theta$) and \texttt{item\_param\_out} (item network $\to a,b,c,d$), are simple linear maps from a learned embedding to the interpretable IRT quantities.

Because these two layers are literally the last linear layer in each branch, feeding directly into the 4PL likelihood, they are natural targets for a last-layer Laplace approximation. We propose to place calibrated uncertainty exactly where the interpretable parameters live, without touching the (nonlinear, harder-to-approximate) embedding networks that precede them.

\section{Method: Last-Layer Laplace Approximation}
\label{sec:method}

\subsection{Background}

The Laplace approximation proposed in \citep{mackay1992evidence} approximates the posterior over a trained network's weights as a Gaussian centered at the maximum-a-posteriori (MAP) $\hat{w}$, with covariance given by the inverse Hessian of the negative log-posterior at $\hat w$, i.e., 
\begin{equation}
p(w \mid \mathcal{D}) \approx \mathcal{N}(\hat{w}, H^{-1}).
\end{equation}
Computing $H$ over all network weights is intractable for anything but tiny networks. Therefore, the \emph{last-layer} variant restricts this to the final linear layer, treating the preceding feature extractor as fixed at its trained value. This framework reduces the problem to (generalized) Bayesian logistic regression on learned features, which is cheap and numerically stable \citep{daxberger2021laplace}.

\subsection{Application to PSN-IRT}
\label{subsec:application}

We apply this to PSN-IRT's two readout layers, using the real 4PL logit: for the ability branch, we hold the answering item's discrimination and difficulty fixed at their MAP values and fit a Laplace posterior over \texttt{student\_ability\_out}, using the true logit $a_{\text{MAP}}(\theta - b_{\text{MAP}})$ against real (model, item, label) triples. Symmetrically, for the item branch, we hold the responding model's ability fixed at its MAP value and fit a Laplace posterior over \texttt{item\_param\_out}. This yields two conditional posteriors, each capturing uncertainty in one side of the model while the other is treated as known, representing a standard plug-in simplification.

\subsection{Calibration validation and a necessary scoping decision}
\label{subsec:calibration}

We validate the posterior's calibration on held-out triplets (model, item, label) using expected calibration error (ECE) and negative log-likelihood (NLL) of the posterior predictive probability. We find that the \emph{ability} branch, sampling $\theta$ alone, is well calibrated (ECE~$\approx 0.05$-$0.07$). On the other hand, the \emph{item} branch, is poorly calibrated when both discrimination $a$ and difficulty $b$ are sampled jointly (ECE~$\approx 0.18$-$0.26$, unaffected by a 10$\times$ increase in fitting data), because $a$ and $b$ are combined multiplicatively in the logit $a(\theta - b)$. Sampling two correlated multiplicative factors compounds variance beyond what a Gaussian last-layer approximation supports. We verify this by holding $a$ fixed at its MAP value and sampling only $b$ restoring calibration to ECE~$\approx 0.08$, matching the ability branch. For a complete report of full calibration, find figures in Appendix~\ref{app:calibration}.

Based on this validation, all downstream results use the following scoped posterior: model ability $\theta$ is treated as fully random (sampled from its Laplace posterior); item difficulty $b$ is treated as random (sampled); item discrimination $a$, guessing rate $c$, and feasibility $d$ are held at their MAP point estimates. We consider this an important methodological finding in its own right: applying last-layer Laplace to models with multiplicatively-interacting output heads requires explicit validation of \emph{which} outputs can jointly carry posterior uncertainty, not merely whether the layer as a whole is amenable to a Gaussian approximation.

\subsection{Posterior-expected Fisher information}

The original Fisher information for the 4PL model is
\begin{equation}
I(\theta) = a^2 \cdot \frac{(P(\theta) - c)^2 (d - P(\theta))^2}{(d-c)^2\, P(\theta)(1-P(\theta))}.
\label{eq:fisher}
\end{equation}
Item selection in \citet{zhou2026lost} uses $I(\hat\theta)$, which is defined as $I(\theta)$ evaluated at point-estimate parameters. Here, we compute the posterior-expected information under our validated, scoped posterior,
\begin{equation}
\bar{I}(\theta) = \mathbb{E}_{p(b \mid \mathcal{D})}\big[I(\theta \mid a_{\text{MAP}}, b, c_{\text{MAP}}, d_{\text{MAP}})\big],
\end{equation}
approximated by Monte Carlo averaging Eq.~\ref{eq:fisher} over posterior samples of $b$. This proposal is the natural Bayesian generalization of Fisher-based item selection used in the computerized adaptive testing literature \citep{vanderlinden1998bayesian,changying1996global}.

\section{Experimental Setup}

We use the PSN-IRT authors' released training code and a 70\% split of their unified response matrix (\texttt{percent\_0.7.csv}), comprising 12 models and 29{,}309 benchmark items across the 11 benchmarks used in the original paper. We train PSN-IRT with the original hyperparameters (Adam, learning rate $3\times10^{-3}$, batch size 512, 30 epochs, embedding dimension 128). We then fit our Laplace posterior as described in Section~\ref{sec:method} and validate calibration on held-out response cells from the same item vocabulary (PSN-IRT's transductive architecture cannot score items outside its trained one-hot vocabulary, so held-out evaluation uses withheld response \emph{cells} rather than withheld items). All posterior sampling is seeded for reproducibility; we verified identical outputs across repeated runs of the full pipeline before reporting the results below (see Appendix~\ref{app:reproducibility}).

Model identities: the released \texttt{combine.csv} does not include a header row or an accompanying model-identifier mapping, so we report results using anonymized model indices (0-11). See in Figure~\ref{fig:theta_ci}.

\section{Results}

\subsection{Pairwise model ability comparisons}

\begin{figure}[t]
\centering
\includegraphics[width=0.85\textwidth]{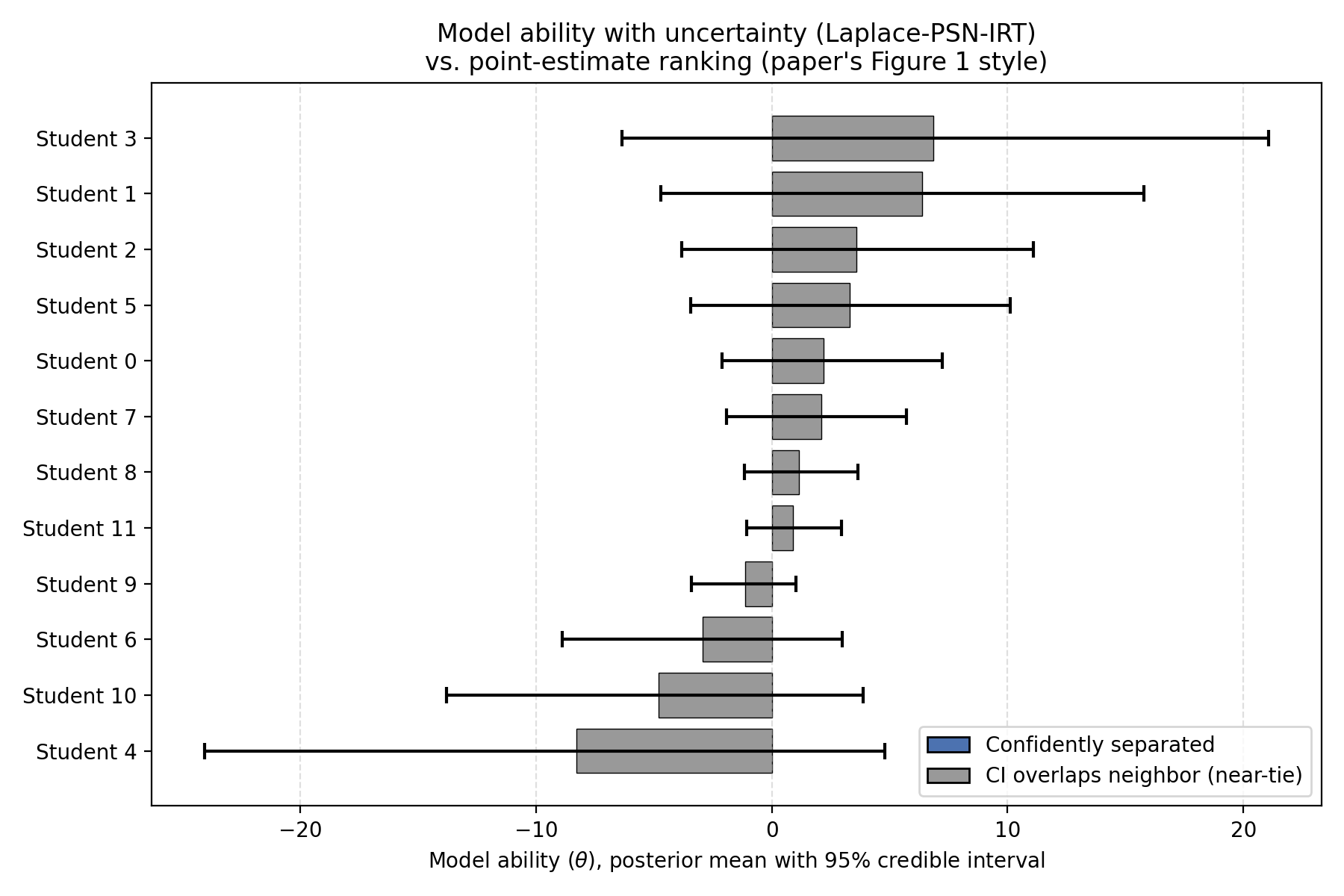}
\caption{Posterior mean model ability ($\theta$) with 95\% credible
intervals, sorted by rank. Anonymized student indices are used throughout
(see Experimental Setup).}
\label{fig:theta_ci}
\end{figure}

In Figure~\ref{fig:theta_ci} we have the posterior mean ability with 95\% credible intervals for all 12 models evaluated. Of the $\binom{12}{2}=66$ pairwise comparisons, only 2 are confidently separated ($P(\theta_i > \theta_j) > 0.95$ or $< 0.05$); the remaining 64 pairs are statistically indistinguishable under this posterior. Note that every adjacent pair in the sorted ranking overlaps in its 95\% credible interval, directly and quantitatively supporting the qualitative ``weak separability'' concern raised by \citet{zhou2026lost}'s Figure 1.

\subsection{Fisher information saturation}
\label{subsec:saturation}

For this part, we compute point-estimate and posterior-expected Fisher information for all items at five reference ability levels (0th, 25th, 50th, 75th, 100th percentile) of the empirical range of model ability. Table~\ref{tab:sat} describes the fraction of items with near-zero ($<10^{-4}$) information under each criterion.

\begin{table}[t]
\centering
\caption{Fraction of items with near-zero Fisher information ($I < 10^{-4}$)
at each reference ability percentile, point-estimate vs. posterior-expected
(scoped posterior, Section~\ref{subsec:calibration}).}
\label{tab:sat}
\begin{tabular}{lccc}
\toprule
Percentile & $\theta_{\text{ref}}$ & \% near-zero (point) & \% near-zero (posterior) \\
\midrule
0th   & $-7.43$ & 70.2\% & 9.3\%  \\
25th  & $-1.64$ & 20.9\% & 0.0\%  \\
50th  & $1.55$  & 44.4\% & 0.0\%  \\
75th  & $3.47$  & 63.4\% & 0.01\% \\
100th & $6.78$  & 88.0\% & 6.5\%  \\
\bottomrule
\end{tabular}
\end{table}

Point-estimate Fisher information collapses for a large, ability-dependent fraction of items and it is worst at the extremes of the ability range, specially where distinguishing the weakest and strongest models matters most. This collapse happens because at a fixed reference ability, far from an item's difficulty, the predicted probability $P(\theta)$ saturates near $c$ or $d$, driving the numerator of Eq.~\ref{eq:fisher} toward zero; posterior averaging over item difficulty largely avoids this, because at least some posterior mass falls away from the saturation boundary.

Point-estimate item rankings are also unstable in direction. The Spearman rank correlation between point-estimate and posterior-expected information changes sign across the ability range we tested (from $+0.28$ at the 25th percentile to $-0.40$ at the 75th percentile), meaning that the two criteria can disagree not merely in magnitude but in which items they would select, depending on which model's ability is used as the reference point.

\subsection{Item selection: recovering a full-benchmark ranking from a subset}
\label{subsec:itemselect}

\citet{zhou2026lost} performs validation of the Fisher-based item selection with an external human-preference reference (Chatbot Arena, aggregated with OpenCompass Arena). Several of the 12 models in the released data are 2023-era open models (e.g. Vicuna-7B-v1.3, Mistral-7B-Instruct-v0.1, Gemma-2B-it) that have since been retired from public arena leaderboards, thus we were unable to reliably reconstruct the archived reference scores used in the original evaluation. Instead, we validate against an internal oracle: the ability ranking obtained from the full item set, and evaluate whether a small selected subset recovers this ranking. We score each subset in two ways: (i) raw mean accuracy on the subset (classical test theory style), and (ii) maximum-likelihood-refit ability on the subset with item parameters held fixed at their trained values (IRT-consistent). We also include a random-selection baseline (mean over 20 draws) for context.

\begin{table}[t]
\centering
\caption{Kendall's $\tau$ between subset-based ranking and the full-item
oracle ranking, for point-estimate vs.\ posterior-expected Fisher-information
item selection, across subset sizes. Bold indicates the better-performing
criterion for that row.}
\label{tab:itemselect}
\begin{tabular}{llcccc}
\toprule
Top-$n$ & Criterion & $\tau$ (accuracy) & $\tau$ ($\theta$-refit) & Random (acc.) & Random ($\theta$-refit) \\
\midrule
50   & Point estimate     & \textbf{0.780} & \textbf{0.846} & 0.788 & 0.871 \\
50   & Posterior expected & 0.717 & 0.737 & 0.776 & 0.864 \\
100  & Point estimate     & 0.681 & 0.561 & 0.792 & 0.891 \\
100  & Posterior expected & \textbf{0.813} & \textbf{0.800} & 0.832 & 0.927 \\
200  & Point estimate     & 0.532 & 0.636 & 0.883 & 0.959 \\
200  & Posterior expected & \textbf{0.772} & \textbf{0.870} & 0.861 & 0.964 \\
400  & Point estimate     & 0.554 & 0.626 & 0.869 & 0.971 \\
400  & Posterior expected & \textbf{0.862} & \textbf{0.870} & 0.861 & 0.983 \\
1000 & Point estimate     & 0.687 & 0.657 & 0.886 & 0.997 \\
1000 & Posterior expected & \textbf{0.870} & \textbf{0.909} & 0.895 & 0.992 \\
\bottomrule
\end{tabular}
\end{table}

Posterior-expected Fisher information outperforms point-estimate Fisher information in 8 of the 10 tested configurations (See Table~\ref{tab:itemselect}) at every item subset size of 100 or larger, under both scoring methods. At the smallest subset size of 50 items, point-estimate Fisher information is modestly ahead under both scoring methods. We interpret this as evidence that posterior averaging over item difficulty is particularly valuable once a subset is large enough to contain a reasonably representative sample of items across the difficulty range. However, at very small subset sizes, a handful of individual point-estimate item-parameter values may happen to align well with the oracle by chance, an effect that averages out as subset size grows. We interpret the 100-1000 item range the more practically relevant regime for efficient benchmarking. Moreover, the original paper's own best point-estimate result was reported at $n=1000$.

As a final transparent note of the results, we remark that neither criterion reliably outperforms random item selection against this internal oracle, at any subset size tested. We attribute this to a structural property of the oracle itself rather than a weakness of Fisher-based selection. The oracle ability ranking is derived from the same trained model and item parameters used to score every candidate subset, recovering it is a comparatively easy statistical task for \emph{any} reasonably sized, non-degenerate subset (informed or random) particularly under $\theta$-refit scoring, where random baselines reach $\tau \approx 0.86$-$0.99$ across the sizes we tested. This self-consistency is precisely what an external reference, such as the Chatbot Arena comparison used in the original paper, avoids. We view reproducing that external comparison as important future work rather than a result we can currently report.

\section{Limitations}

\textbf{Item-branch posterior is scoped, not full.} As described in Section~\ref{subsec:calibration}, our validated posterior treats item discrimination, guessing rate, and feasibility as fixed point estimates. In this paper only difficulty and model ability carry posterior uncertainty. This is a direct consequence of a calibration failure mode we identify and validate.

\textbf{No external reference for item selection.} As discussed in Section~\ref{subsec:itemselect}, we validate item selection against an internal, model-derived oracle rather than the external human-preference reference used by \citet{zhou2026lost}, due to the unavailability of archived arena scores for several legacy models in the released dataset.

\textbf{Anonymized model identities.} The released response matrix does not include a model-ID mapping; we report results using anonymized indices. This does not affect the validity of our methodological contribution, which concerns the uncertainty-quantification method rather than any specific model's identity, but it limits the direct interpretability of Figure~\ref{fig:theta_ci} for readers wishing to identify specific models.

\textbf{Plug-in conditional posteriors.} Our ability and difficulty posteriors are each fit holding the other side's parameters fixed at their MAP value (Section~\ref{subsec:application}), rather than as a single joint posterior over both branches simultaneously. In the beginning, we attempted a joint last-layer Laplace fit over both branches together and found it performed \emph{worse} than the two scoped conditional fits (higher ECE, unaffected by additional fitting data). We report this as a negative result in Appendix~\ref{app:calibration} rather than pursuing it further, but note it as an open direction for future work on multi-branch Laplace approximations.

\textbf{Item-selection advantage is size-dependent.} As reported in Section~\ref{subsec:itemselect}, posterior-expected Fisher information does not uniformly outperform point-estimate Fisher information because at the smallest subset size we tested (50 items), point estimate is modestly ahead. We report this transparently rather than selectively presenting only the subset sizes at which our method wins.

\section{Related Work}

\citet{zhou2026lost} is the direct basis for this work and we summarize it in Section 2. The Laplace approximation for Bayesian deep learning follows \citet{mackay1992evidence}, with the modern last-layer treatment and practical library (\texttt{laplace-torch}) from \citet{daxberger2021laplace}. \citet{vanderlinden1998bayesian} Motivated the Bayesian item-selection criteria for adaptive testing, serving as inspiration for our posterior-expected Fisher information. A related global-information alternative for non-Gaussian posteriors is given by \citet{changying1996global}. In \citet{maiapolo2024tinybenchmarks} it is similarly pursued the efficient benchmark construction via IRT-based item selection, using point estimates. \citet{madaan2024quantifying} and \citet{miller2024adding} independently raise the need for variance/error-bar reporting in LLM benchmark evaluation, a concern our posterior-based results directly address.

\section{Conclusion}

In this article, we show that a lightweight, post-hoc last-layer Laplace approximation, applied carefully, with explicit validation of which parameters can jointly carry posterior uncertainty, yields calibrated model-ability and item-difficulty posteriors for PSN-IRT without retraining. The proposed posteriors reveal an important conclusion: the large majority of pairwise model comparisons on a standard LLM leaderboard are not statistically distinguishable, that point-estimate Fisher information used for efficient benchmark construction is systematically fragile due to saturation, and that posterior-expected Fisher information is a better item-selection criterion at practically relevant subset sizes. We hope this motivates wider adoption of calibrated uncertainty quantification in neural IRT models of LLM evaluation.

\bibliographystyle{plainnat}

\appendix

\section{Calibration validation details}
\label{app:calibration}

We validate posterior calibration on 5{,}000 held-out (model, item, label) triples via Expected Calibration Error (ECE) and negative log-likelihood (NLL), under six configurations: the MAP model alone (no posterior sampling); the ability branch alone ($\theta$ sampled, item parameters fixed at MAP); the item branch with both discrimination and difficulty sampled jointly; the item branch with difficulty sampled alone (discrimination fixed at MAP); a single joint Laplace fit spanning both branches simultaneously (Section~\ref{subsec:application}); and the full two-branch model with both branches' samples combined into a single predictive probability. Table~\ref{tab:calibration} summarizes.

\begin{table}[h]
\centering
\caption{Calibration (Expected Calibration Error and negative log-likelihood)
under different posterior-sampling configurations, on 5{,}000 held-out
response cells.}
\label{tab:calibration}
\begin{tabular}{lcc}
\toprule
Configuration & ECE & NLL \\
\midrule
MAP (no posterior sampling) & 0.033 & 0.285 \\
Ability branch alone ($\theta$ sampled) & 0.047 & 0.258 \\
Item branch, $a$ and $b$ jointly sampled & 0.182 & 0.383 \\
Item branch, $b$ only sampled ($a$ fixed) & 0.081 & 0.293 \\
Joint single Laplace fit (both branches) & 0.249 & 0.467 \\
Combined two-branch predictive (both branches' samples) & 0.189 & 0.390 \\
\bottomrule
\end{tabular}
\end{table}

The ability branch alone and the item branch's difficulty-only configuration achieve comparable, well-calibrated ECE ($\approx 0.05$ and $\approx 0.08$ respectively), while every configuration that jointly samples discrimination and difficulty -- whether combined across branches, fit jointly in a single Laplace approximation, or isolated to the item branch alone -- is substantially miscalibrated (ECE $0.18$--$0.25$). This motivates the scoped posterior used throughout the paper (Section~\ref{subsec:calibration}).

\begin{figure}[h]
\centering
\includegraphics[width=0.55\textwidth]{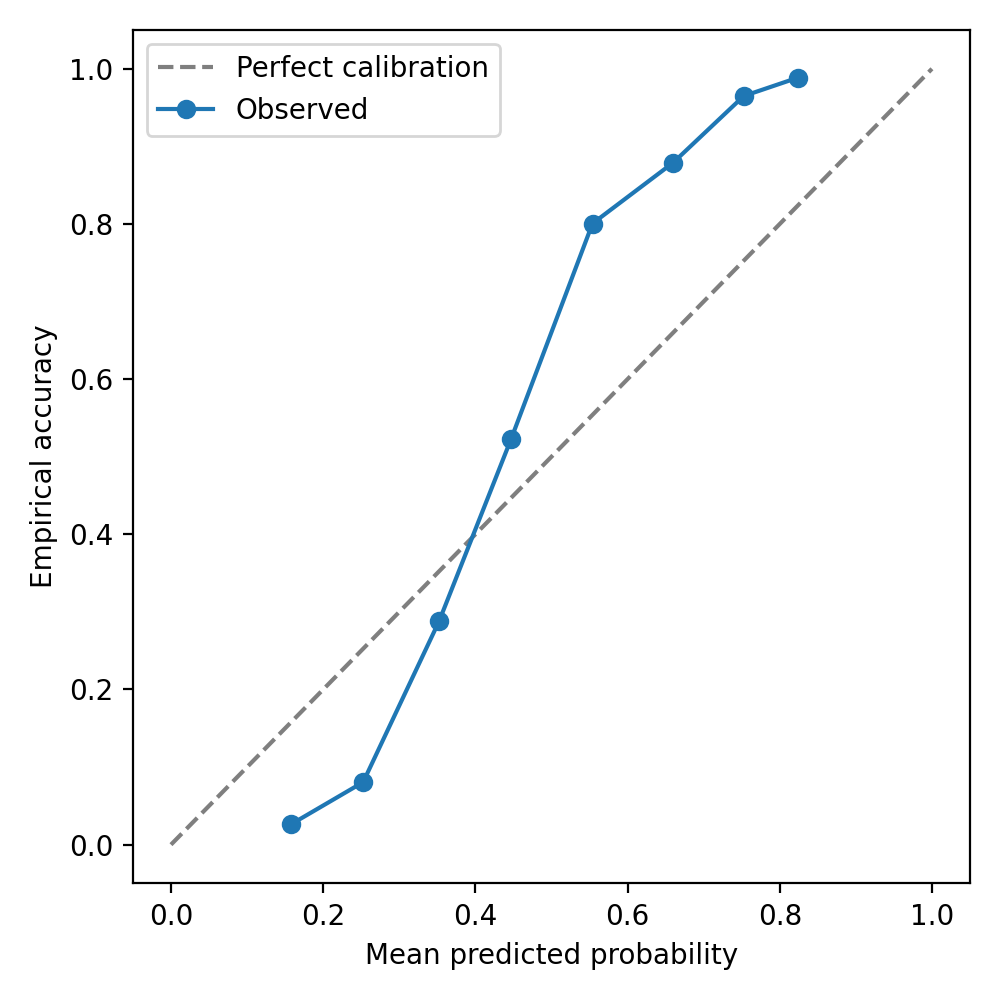}
\caption{Reliability diagram for the combined two-branch posterior predictive
(both branches' samples, corresponding to the ``Combined two-branch
predictive'' row of Table~\ref{tab:calibration}), computed on the same
5{,}000 held-out response cells. The dashed line indicates perfect
calibration. The observed curve lies below the diagonal at low predicted
probabilities and above it at high predicted probabilities, the signature
of underconfident, compressed predictions -- consistent with the
flattening-toward-0.5 behavior discussed in Section~\ref{subsec:calibration}
and Appendix~\ref{app:calibration}.}
\label{fig:reliability}
\end{figure}

\section{Reproducibility}
\label{app:reproducibility}

An earlier version of our pipeline exhibited run-to-run variation in posterior-sampled quantities despite fixed random seeds for Python's \texttt{random} and PyTorch's random number generators; we traced this to the posterior-sampling routine in the Laplace approximation library relying on NumPy's random number generator, which was not separately seeded. Seeding NumPy's generator alongside Python's and PyTorch's resolved this: we verified byte-identical (MD5-matched) outputs for all posterior-dependent result files across repeated, independent runs of the full pipeline before reporting any numbers in this paper. All numbers reported in Sections~\ref{subsec:saturation}--\ref{subsec:itemselect} and Figure~\ref{fig:theta_ci} reflect this verified, reproducible configuration.

Code and a full reproduction guide (\texttt{REPRODUCE.md}) are available at \url{https://github.com/JFMandujanoR/Laplace-PSN-IRT/tree/jfmr_laplace}. All experiments use the response matrix and training code released by \citet{zhou2026lost} at \url{https://github.com/Joe-Hall-Lee/PSN-IRT}.

\end{document}